\documentclass[showpacs,preprintnumbers]{revtex4-2}
\usepackage{amssymb}
\usepackage{graphicx}
\usepackage{color}

\def\bc{\begin{center}}
\def\ec{\end{center}}

\def\beq{\begin{equation}}
\def\eeq{\end{equation}}

\usepackage{graphicx}
\usepackage{caption}
\usepackage{subcaption}
\usepackage[T1]{fontenc}
\usepackage[utf8]{inputenc}
\usepackage[english]{babel}
\usepackage[hidelinks]{hyperref}
\usepackage{indentfirst}
\usepackage{amssymb}
\usepackage{amsmath}
\usepackage{mathrsfs}
\usepackage{braket}
\usepackage{float}
\usepackage{color}

\begin{document}
\title{Quantum entanglement between excitons in two-dimensional materials}

\author{Gabriel P. Martins$^{1,2}$, Oleg L. Berman$^{1,2}$, Godfrey Gumbs$^{2,3}$ and Yurii E. Lozovik$^{4,5}$}
\affiliation{\mbox{$^{1}$Physics Department, New York City College
of Technology, The City University of New York,} \\
Brooklyn, NY 11201, USA \\
\mbox{$^{2}$The Graduate School and University Center, The
City University of New York,} \\
New York, NY 10016, USA\\
\mbox{$^{3}$Department of Physics and Astronomy, Hunter College of The City University of New York
City University of New York,} \\
New York, NY 10065, USA\\
\mbox{$^{4}$Institute of Spectroscopy, Russian Academy of Sciences, Troitsk, Moscow, Russia 142190}
\mbox{$^{5}$Research University Higher School of Economics, Moscow, Russia 101000}
}
\date{\today}

\begin{abstract}

The quantum entanglement between two excitons  in  two-dimensional materials, embedded in an optical microcavity, was investigated. The energy eigenstates of a Jaynes-Cummings like Hamiltonian for two qubits coupled to a single cavity mode have been calculated. The quantum entanglement between  such states was estimated by calculating the concurrence between two qubits in each of these eigenstates. According to the results of our calculations, if the system is allowed to decay only through the emission of cavity photons at low temperatures, there is a maximally entangled eigenstate, protected from decay. We demonstrated that the existence of such a state results in the counter-intuitive conclusion  that, for some initial states of the system, the fact that the cavity is leaky can actually lead to an increase in the average concurrence on the timescales of the average photonic lifetime. In addition, we calculated the time evolution of the concurrence between a pair of excitons in a strained graphene monolayer.

\end{abstract}

\pacs{PACs}
\maketitle

\section{Introduction}
\label{sec1}

{Modern quantum technologies and the prospects for  their development are based on the use of quantum two-level (2D) systems as   qubits.   A number of physical realizations of qubits in systems with discrete spectra have been discussed. These include  ultra cold ions and atoms in traps, impurities in diamond, and various types of superconducting qubits. Spatially extended semiconductors and new 2D materials, as well as transition metal dichalcogenides (TMDCs) all have a band structure, However, in the  presence of transverse magnetic field,  a 2D system  has a discrete spectrum consisting of degenerate Landau levels. When electrons pass from filled Landau levels to unfilled excited electron states, then  due to the Coulomb attraction between electrons and holes, they form 2D magnetoexcitons,   the energy of which depends continuously on the integral of motion in a magnetic field, i.e.,  the magnetic momentum. This integral of motion is a consequence of the invariance of the system with respect to simultaneous translation and gauge transformation (see Refs.~\cite{lozovik1997magnetoexcitons,lozovik1997magnetoexciton2}). As a result of the continuous dependence of the magnetoexciton energy on the magnetic  pulse, the full spectrum of the system is not discrete, but consists  of bands (see Ref.~\cite{lerner1980mott}).  

\medskip
\par

Low-dimensional 2D materials have been the subject of an enormous amount of scrutiny within the past few decades \cite{ando1982electronic,bychkov1984oscillatory, beenakker1991quantum}, but, ever since the already historic work where Novoselov, Geim and their collaborators managed to obtain both monolayer and bilayer samples of graphene  in 2004 \cite{novoselov2004electric}, interest in the field has  increased substantially by the exciting new physical frontiers it opened. Soon thereafter, a great number of researchers began to investigate the properties of graphene \cite{mccann2006landau,guinea2010energy,guinea2010generating}  as well as other  2D materials that quickly followed. All those studies have been very well summarized in a variety of review papers, of which one of the most well-known was published in 2007 by Katsnelson \cite{katsnelson2007graphene}.     Fundamentally new properties of 2D systems are exhibited in pseudomagnetic fields, which arise, in particular, upon deformation of graphene (see Refs.~\cite{guinea2010energy,guinea2010generating,amorim2016novel,guinea2008midgap}).   One of the many interesting properties of  2D materials is that it is the perfect environment for the development of magnetoexcitons \cite{lerner1980mott}, the pseudoparticle consisting of the bound states of a negatively charged electron and a positively charged hole.  There are two types of excitons that can appear on 2D materials. These are direct excitons, when the electron and the hole are on the same layer, and indirect excitons,   when they are on different layers.

\medskip
\par

 An interesting discovery in recent years has been the fact that  when mechanical strain is applied to a monolayer of  graphene, a pseudo-magnetic field   is generated within that sheet \cite{levy2010strain}. For this system, excitons, formed by an electron and a hole in different valleys,    have dispersionless Landau levels \cite{berman2020strain}, whereas excitons in a uniform external magnetic field display   a non-trivial energy dispersion relation \cite{lerner1980mott}. This gives rise to well defined energy gaps and might be employed as qubits,  if one considers only the ground state and first excited state of each exciton.    Our focus in this paper will be cases where we investigate the entanglement between two such qubits induced by their interaction with a cavity mode. Another way to restrict excitons to discrete energy levels in 2D          is to trap them in harmonic potentials. One way to do this is through pinning the 2D material with a thin needle \cite{negoita1999stretching},         or to apply a potential difference between two layers of the material \cite{voros2006trapping}.  These methods to  achieve discrete energy levels for excitons can be applied  to various 2D materials, including    a class of monolayers of direct band-gap materials, namely, TMDCs \cite{Kormanyos,Glazov}.

\medskip
\par

 Another important property of graphene is the chirality of its two independent valleys. This leads to the fact that circularly polarized light is absorbed in different valleys, depending on the sign of the circular  polarization. In addition, a consequence of the chirality of the valleys is that the pseudomagnetic field arising during deformation has opposite directions in different valleys. As a result of this, the appearance of a pseudomagnetic field during deformation does not contradict the absence of violation of invariance with respect to time reversal in the system in the absence of a real magnetic field.

\medskip
\par
When graphene is pumped by plane-polarized photons, the absorption of photons occurs in both valleys, accompanied by the appearance of electrons and holes in them. As a result of Coulomb attraction, these electrons and holes form excitons. Moreover, electrons can bind both with holes from their own valley (intravalley excitons) and with holes from another valley (intervalley excitons).

\medskip
\par
Since the pseudomagnetic field acts in the same way on an electron and a hole, and the pseudomagnetic fields in different valleys have the opposite direction (opposite sign), the spectrum of intervalley pseudomagnetoexcitons, as can be easily shown, turns out to be discrete, in contrast to magnetoexcitons in real magnetic fields. Therefore, intervalley excitons can be used as quantum elements for quantum technologies. A pair of intervalley excitons can be located either in the same pair of graphene valleys, or in different pairs of valleys, and the electrons in each pair of valleys have two choices for holes. Therefore, there are four states for intervalley excitons, which have the same energies and other properties due to the symmetry between the valleys. Thus, intervalley excitons are actually ququarts, which can also be used as intermediate elements in quantum technologies  \cite{kiktenko2020scalable}. Note that the properties of the considered ququarts can be controlled using an external real magnetic field.that removes degeneracy and splits levels. This can be used in quantum technologies.

\medskip
\par

Quantum entanglement is a degree of bonding which quantum subsystems might have that has no classical counterpart. Since it cannot be created locally (by acting on a single subsystem), but it can be transferred from one subsystem to another, it is usually treated as an important resource in quantum information and quantum computation \cite{wootters1998quantum}, and, like any useful resource, quantifying it has become an important challenge. Many different ways of measuring the entanglement of a quantum system have been suggested \cite{bennett1996mixed,vedral1997quantifying, donald1999continuity,wootters2001entanglement}.   One of those is  Concurrence \cite{wootters2001entanglement}, a measure of entanglement between a pair of two-leveled quantum systems, or simply, two qubits.

\medskip
\par
Recently,  the conditional concurrence created by the dynamical Lamb effect of two qubits within a microcavity which changes its frequency suddenly was  calculated \cite{berman2016quantum}. In this paper, we follow the same protocol, and find the concurrence created by two pseudomagnetic excitons on a graphene sheet under strain embedded in a microcavity interacting with one of its modes in the rotating wave approximation (RWA).

\medskip
\par

The rest of the paper is organized as follows: In Sec. \ \ref{sec2},  we define the system  under investigation, which consists of a pair of excitons on
 a graphene sheet under strain inside an optical microcavity.  In Sec. \  \ref{sec3},    we present and solve a model Hamiltonian for the system. ection \  \ref{sec4} is devoted  to a     study of the entanglement created by the dynamics through calculating the concurrence between the excitons 
  We  study in Sec. \ \ref{sec5} what happens when the system is subject to dissipation. In Sec. \ \ref{sec6},   we investigate the time evolution of the entanglement between those two excitons. This is accomplished  by calculating the time evolution of the concurrence.   In Sec. \ref{sec7}, we apply our model to the case of excitons on a graphene sheet under strain and explain how to generalize the result   to other 2D   materials. The possible physical realizations for the system under consideration are   analyzed in  Sec.~\ref{real}.    In Sec. \ref{sec8},  we present a summary of this   work.

\section{Excitons on a Graphene Sheet Under Strain}
\label{sec2}

It has been shown that particles on a graphene sheet that is subjected to strain obey a Hamiltonian which is equivalent to one describing the effects of magnetic field \cite{levy2010strain}. The effect of these so-called pseudomagnetic fields differs from real magnetic fields by the fact that they are charge-independent, affecting positively and negatively charged particles the same way. Strain-induced pseudomagnetic fields in graphene, like external magnetic fields, provide a favorable environment for the formation of excitons, the quasiparticles formed by the bound state of electrons and holes.  Within this environment, the resulting excitons, formed by an electron and a hole in different valleys, have a non-dispersive discrete energy spectrum \cite{berman2020strain}.   In contrast, excitons in an applied magnetic field have discrete energy levels each of which is dispersive \cite{lerner1980mott}. The energy eigenvalues for the  pseudomagnetoexcitons (PMEs) have been calculated \cite{berman2020strain} and  are given by

\begin{equation}
E_{n,\tilde{n},\tilde{m}} =\varepsilon_{0n,\tilde{n}}+E^\prime_{\tilde{n},\tilde{m}}\label{etot} \  ,
\end{equation}
where, for the simplest case in which the electron and hole masses are the same,

\begin{equation}
\varepsilon_{0n,\tilde{n}} = \hbar\omega_c(n+\tilde{n}+1)\label{ezero}.
\end{equation}
The quantum numbers $n$ and $\tilde{n}$
quantify  the cyclotron motion of the center of mass and relative  motion, respectively. The term $E^\prime_{\tilde{n},\tilde{m}}$ represents the excitonic binding energy which depends on a new quantum number $\tilde{m}$. Calculated values for the excitonic binding energies $E^\prime_{\tilde{n},\tilde{m}}$ are presented  in Ref.\  \cite{berman2020strain}. In Eq.\   (\ref{ezero}), the cyclotron frequency $\omega_c$ is given by $\omega_c = \dfrac{B}{m}$, where $B$ is the intensity of the pseudomagnetic field ($\frac{B}{e}$ has units of Teslas) and $m$ is the electronic mass.

\medskip
\par

Since they present a discrete set of energy eigenstates, under  suitable conditions, PMEs can be treated as qubits. Here, we consider a system consisting of two such qubits that are coupled to a single cavity mode and study the quantum entanglement between them.

\begin{figure}[H]
\begin{center}

\includegraphics[scale=0.3]{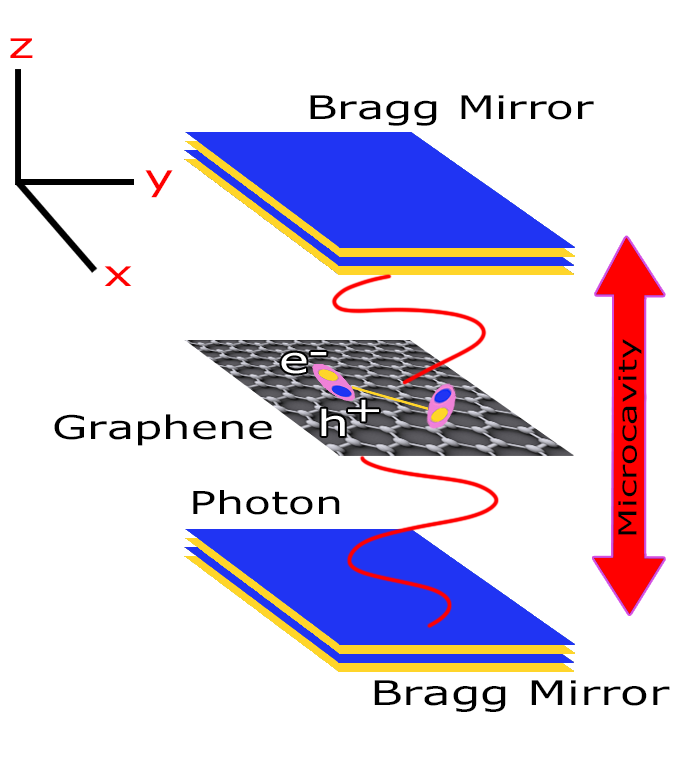}
\caption{System studied throughout the paper, consisting of 2 excitons on a strained sheet of graphene interacting with a single mode inside an optical microcavity.}\label{schem}
\end{center}
\end{figure}

From this point forward, we consider a system consisting of two such PMEs, on a strained graphene sheet inside an optical microcavity (Fig. \ref{schem}). We consider the exciton-exciton interaction to be neglectable and the excitons do not interact directly with one another, but both are in contact with the same cavity mode and are allowed to become entangled by such.

\section{A JC-like model for two qubits in a microcavity}
\label{sec3}

Qubits form the basics for quantum computing, and quantum entanglement is the key resource present in quantum computing that makes it such a powerful tool. Under the right circumstances, we can consider excitons on a graphene sheet under strain to be qubits. Under these conditions, we applied a Jaynes-Cummings-like model to study the quantum entanglement created between two qubits that are coupled to a single cavity mode.
\medskip
\par

Let us consider a system consisting of two qubits interacting with a cavity mode with Hamiltonian

\begin{equation}
\hat{H}=\hat{H}_0+\hat{V^\prime}_{RWA}\label{hami}
\end{equation}
with  the unperturbed Hamiltonian $\hat{H}_0$  given by

\begin{equation}
\hat{H}_0=\hbar\left(\sum_{j=1}^2\omega_0\ket{e_j}\bra{e_j}+\omega_k\hat{a}^\dagger\hat{a}\right)\label{H0} \   ,
\end{equation}
where, in this notation,  $\ket{e_j}$ is  the excited state of qubit $j$, $\hbar\omega_0$ is the energy gap of the qubits, $\omega_k$ is the mode frequency and $\hat{a}^\dagger$ and $\hat{a}$ are the creation and annihilation operators for photons in the cavity.   Here, we consider the excited state for the qubit, $\ket{e}$, to consist of an exciton in its first excited state, and $\ket{g}$ to consist of that same exciton in the ground state. Also, for simplicity, in Eq.\  (\ref{H0}), we relabel the excitonic energies so that the energy of an exciton in the ground state is set equal to 0. The energy gap $\hbar\omega_0$ comes directly from Eq.\ (\ref{etot}) and is equal to $\hbar\omega_0 = E^\prime_{0,1}-E^\prime_{0,0}$.    Therefore, the eigenstates of $\hat{H_0}$ are $\ket{n;ij}$, having  energy $E_{nij}$ given by

\begin{equation}
E_{nij} = \hbar [\omega_0(i + j) + n\omega_k]
\end{equation}
where $i = 1$ ($j = 1$) if the first (second) qubit is in the excited state $\ket{e}$ and $i = 0$ ($j = 0$) if the first (second) qubit is in the ground state $\ket{g}$.

\medskip
\par

The interaction Hamiltonian on the rotating wave approximation (RWA), $\hat{V^\prime}_{RWA}$ is given by \citep{ballentine2014quantum}

\begin{equation}
\hat{V^\prime}_{RWA} = \hbar\lambda\sum_{j=1}^2\left(\hat{\sigma}_j^+\hat{a}+\hat{\sigma}_j^-\hat{a}^\dagger\right)
\end{equation}
where $\hat{\sigma}_j^\pm$ are the creation (+) and annihilation (-) operators for the qubit $j$ and $\lambda$ is a coupling constant.

\medskip
\par

This Hamiltonian is similar to that for     the Jaynes-Cummings model \cite{jaynes1963comparison,shore1993jaynes}, except that it is for two qubits in one cavity mode. Since the qubits interact with the mode according to the RWA, the destruction (creation) of photons is determined by the creation (annihilation) of qubit excitations. Therefore,  as for the Jaynes-Cummings model, we can treat each manifold with fixed number of excitation, meaning the sum of number of the number of photonsand the number of qubit excitations, individually. The $n$-th manifold is composed of the states with $n$ total excitations, namely $\ket{n;00}\equiv\ket{0_n}$, which is the state with $n$ cavity photons and no qubit excitations; $\ket{n-1;01}\equiv\ket{1_n}$, the state with $n-1$ cavity photons and with  qubit ``1" in the excited state $\ket{e}$; $\ket{n-1;10}\equiv\ket{2_n}$,  the state with $n-1$ cavity photons and with the qubit ``2" in the excited state $\ket{e}$ and $\ket{n-2;11}\equiv \ket{3_n}$, the state with $n-2$ total excitations and both qubits in the excited state. It is important to note that each manifold is 4-dimensional with exception to the $n=1$, which is 3-dimensional and to the $n=0$ which is composed only with  the ground  state, $\ket{0;00}$. The effective Hamiltonian, $\hat{H}_n$, on the $n$-th manifold appears directly from Eq.\ (\ref{hami}) and is equal to

\begin{equation}
H_1 = \hbar \begin{pmatrix}
\omega_k &\lambda&\lambda\\
\lambda & \omega_0 &0 \\
\lambda & 0 & \omega_0
\end{pmatrix},\label{H1}
\end{equation}
for $n=1$ and

\begin{equation}
H_n = \hbar \begin{pmatrix}
n\omega_k &\lambda\sqrt{n}&\lambda\sqrt{n}& 0\\
\lambda\sqrt{n} & (n-1)\omega_k+\omega_0 &0 & \lambda\sqrt{n-1}\\
\lambda\sqrt{n} & 0 & (n-1)\omega_k+\omega_0 &\lambda\sqrt{n-1} \\
0 & \lambda\sqrt{n-1}& \lambda\sqrt{n-1}&(n-2)\omega_k+2\omega_0
\end{pmatrix},\label{Hn}
\end{equation}
when $n=2,3,\cdots$. For the ground state, $H\ket{0;00}=0$. One can find in a straightforward way the energy eigenstates for the Hamiltonian in (\ref{H1}). The eigenvalues of (\ref{H1}) are

\begin{eqnarray}
\epsilon_0 &=& \omega_0\label{l11}
\nonumber\\
\epsilon_\pm &=& \dfrac{1}{2}\left((\omega_0+\omega_k)\pm\sqrt{(\omega_0-\omega_k)^2+{8\lambda^2}}\right),\label{l12}
\end{eqnarray}
with eigenvectors

\begin{eqnarray}
\ket{\psi_0} &=& \dfrac{1}{\sqrt{2}}\left(\ket{0;01}-\ket{0;10}\right)\label{vec11}
\nonumber\\
\ket{\psi_\pm} &=& \dfrac{1}{\sqrt{2+a_\pm^2}}\left(a_\pm\ket{1;00}-\ket{0;01}-\ket{0;10}\right),\label{vect12}
\end{eqnarray}
where $a_\pm$ is given by

\begin{equation}
a_\pm = \dfrac{1}{2\lambda}\left[(\omega_0-\omega_k)\pm\sqrt{(\omega_0-\omega_k)^2+{8\lambda^2}}\right].\label{apm}
\end{equation}

\medskip
\par

The eigenvalues and eigenvectors of (\ref{Hn}), however, are much more difficult to obtain.   However, this problem is solved  when we consider the resonance case where the cavity frequency is the same as the qubit transition frequency, $\omega_0=\omega_k \equiv \omega$. In this case, Eqs.\ (\ref{H1}) and (\ref{Hn}) become

\begin{eqnarray}
H_1 &=& \hbar \begin{pmatrix}
\omega &\lambda&\lambda\\
\lambda & \omega &0 \\
\lambda & 0 & \omega
\end{pmatrix}\label{H1res}
\\
H_n &=& \hbar \begin{pmatrix}
n\omega &\lambda\sqrt{n}&\lambda\sqrt{n}& 0\\
\lambda\sqrt{n} & n\omega &0 & \lambda\sqrt{n-1}\\
\lambda\sqrt{n} & 0 & n\omega &\lambda\sqrt{n-1} \\
0 & \lambda\sqrt{n-1}& \lambda\sqrt{n-1}&n\omega
\end{pmatrix}\label{Hnres}. \  .
\end{eqnarray}
The eigenvalues of (\ref{H1res}) are

\begin{eqnarray}
\epsilon_0 &=& \omega
\nonumber\\
\epsilon_\pm &=& \omega\pm \sqrt{2}\lambda,
\end{eqnarray}
with eigenvectors

\begin{eqnarray}
\ket{\psi_0} &=& \dfrac{1}{\sqrt{2}}\left(\ket{0;01}-\ket{0;10}\right)\label{eg11}
\\
\ket{\psi_\pm} &=& \dfrac{1}{2}\left(\pm\sqrt{2}\ket{1;00}-\ket{0;01}-\ket{0;10}\right)\label{eg12},
\end{eqnarray}
which are  simply Eqs.\ (\ref{l11}-\ref{vect12}) at resonance.

\medskip
\par

The eigenvalues of Eq.\ (\ref{Hnres}) are given by

\begin{eqnarray}
{\epsilon_n}_0 &=& n\omega
\\
{\epsilon_n}_\pm &=& n\omega\pm\sqrt{2(2n-1)}\lambda  \  ,
\end{eqnarray}
where the first eigenvalue is doubly degenerate. The corresponding eigenvectors are as follows

\begin{eqnarray}
\ket{{\psi_n}_{0,1}} &=& \dfrac{1}{\sqrt{2}}(\ket{n-1;01}-\ket{n-1;10})\label{egn1}
\\
\ket{{\psi_n}_{0,2}} &=&\dfrac{1}{\sqrt{2n-1}}\left(\sqrt{n-1}\ket{n;00}-\sqrt{n}\ket{n-2;11}\right)\label{psi_2}
\\
\ket{{\psi_n}_\pm} &=& \dfrac{1}{\sqrt{8n-4}}\left[\sqrt{2n}\ket{n;00}+\sqrt{2(n-1)}\ket{n-2;11}\pm\sqrt{2n-1}\left(\ket{n-1;01}+\ket{n-1;10}\right)\right]\  .
\label{egnl}
\end{eqnarray}
Here,  $\ket{{\Psi_n}_{0,1}}$ and $\ket{{\Psi_n}_{0,2}}$ are the two degenerate eigenstates with eigenvalue $\epsilon_{n0}$ and $\ket{{\psi_n}_\pm}$ are the eigenstates corresponding to the eigenvalue $\epsilon\pm$.    It is interesting to note that the eigenvalues and eigenvectors of Eqs.\ (\ref{H1res}) and (\ref{Hnres}) do agree with each other for the case $n=1$,  with the exception of the eigenvector $\ket{{\psi_n}_{0,2}}$, which, for $n=1$, contains a vector that does not exist in reality, which means that a state with 1 photon, as can be readily seen from Eq.\  (\ref{psi_2}).

\section{Concurrence for Energy Eigenstates}
\label{sec4}

Now that we have the energy eigenstates for the Hamiltonian, we can determine the quantum entanglement for each of them. The most used measure for quantum entanglement between two qubits is the concurrence \cite{wootters1998quantum} which is what we employ here. We can use Eq.\ (\ref{egn1}-\ref{egnl}) to determine the concurrence between a pair of qubits in energy eigenstates for all $n>1$ and Eqs.\ (\ref{eg11}) and (\ref{eg12}) to calculate the concurrence for all energy eigenstates corresponding to $n=1$. The concurrence for a pair of qubits in state $\ket{\psi}=a\ket{00}+b\ket{11}+c\ket{10}+d\ket{01}$, where $\braket{\psi|\psi}=1$ is given by \cite{wootters1998quantum}

\begin{equation}
C(\psi) = 2|ab-cd|  \  .
\label{conc}
\end{equation}
For a mixed state with density matrix $\rho$, the concurrence of the system is defined by \cite{wootters2001entanglement}

\begin{equation}
C = \max \{0,\lambda_1-\lambda_2-\lambda_3-\lambda_4\}  \  ,
\label{concm}
\end{equation}
where $\lambda_i$ are the eigenvalues in descending order of $\tilde{\rho}\rho$. The matrix $\tilde{\rho}$ is the result of applying the spin-flip operator to $\rho$ so that

\begin{equation}
\tilde{\rho} = (\sigma_y\otimes\sigma_y)\rho^\ast (\sigma_y\otimes\sigma_y)  \  .
\end{equation}
By substituting Eq.\  (\ref{egn1}-\ref{egnl}) in Eq.\ (\ref{conc}), we  obtain the concurrence for the $n>1$ energy eigenstates to be

\begin{eqnarray}
C({\psi_n}_{0,1}) =& 1 &n\geq 1
\nonumber\\
C({\psi_n}_{0,2}) =&\dfrac{2n\sqrt{1-\dfrac{1}{n^2}}}{2n-1} &; n\geq 2
\nonumber\\
C({\psi_n}_\pm) =&\dfrac{\left|1-2n\left(1-\sqrt{1-\dfrac{1}{n^2}}\right)\right|}{4n-2} &; n\geq 1   \  .
\end{eqnarray}
Also, for $n=1$, we substitute Eqs.\  (\ref{eg11}) and (\ref{eg12}) into Eq.\ (\ref{conc}) to find

\begin{eqnarray}
C(\psi_0) &=& 1
\nonumber\\
C(\psi_\pm) &=& \dfrac{1}{2}  \  .
\end{eqnarray}
It is interesting to note that, in the case when there is resonance, the concurrence does not depend on any of the system parameters (i.e., the energy gap $\omega$ and the coupling strength $\lambda$). The concurrence of an eigenstate of the Hamiltonian which is in a superposition of $\ket{{\psi_n}_{0,1}}$ and $\ket{{\psi_n}_{0,2}}$ such as $\ket{\Psi} = \dfrac{1}{\sqrt{|a|^2+|b|^2}}\left(a\ket{{\psi_n}_{0,1}}+b\ket{{\psi_n}_{0,2}}\right)$ is given by

\begin{equation}
C(\Psi) = \dfrac{1}{|a|^2+|b|^2}\left| a^2-b^2\dfrac{2n\sqrt{1-\dfrac{1}{n^2}}}{2n-1}\right|
\end{equation}
which can  vary continuously    between 0  and 1.
\medskip
\par

If the system is off resonance for the states represented by Eq.\ (\ref{vec11}) or (\ref{vect12}), the concurrence will be equal to

\begin{eqnarray}
C(\psi_0) = 1
\nonumber\\
C(\psi_\pm) = \dfrac{2}{2+a_\pm^2},
\end{eqnarray}
where $a_\pm$ is given by Eq.\ (\ref{apm}).   We see that in this case the concurrence depends only on the dephasing $(\omega_0-\omega_k)$ and on the coupling strength between the qubits and a cavity mode photon, $\lambda$, but not on the individual frequencies $\omega_0$ and $\omega_k$. It is reasonable to expect that the same will be true for $n>1$.

\section{Dissipative System}
\label{sec5}

So far, we have considered ideal systems that do not interact with the environment. For those systems, the quantum entanglement is preserved, not decaying with time. However, real  systems  experience various forms of dissipation that end up destroying the entanglement between the qubits. In this section, we  study the dynamics of the same system as described above , but now in the presence of dissipation, in order to see how the concurrence (and, with it, the quantum entanglement) evolves in time.
\medskip
\par

From this point onward, we assume that this system is not isolated and, therefore, it suffers dissipation. When a system experiences any form of dissipation, its time evolution ceases to obey   Schr\"odinger's equation. The most common forms of dissipation force the system to evolve under a master equation such as \cite{nielsen2002quantum}

\begin{equation}
\dot{\rho} = -i[\hat{H},\rho]+\kappa\mathcal{L}(\hat{a})\rho +\sum_{j=1}^2\left(\gamma\mathcal{L}(\hat{\sigma}_j^-)\rho+\gamma_\phi\mathcal{L}\left(\hat{\sigma}_j^{(3)}\right)\rho\right), \label{master}
\end{equation}
where $\rho$ is the system's density matrix, the Lindblad operators $\mathcal{L}$ are defined as \cite{nielsen2002quantum}

\begin{equation}
\mathcal{L}(\hat{A})\rho = \hat{A}\rho\hat{A}^\dagger -\dfrac{1}{2}\left(\hat{A}^\dagger\hat{A}\rho+\rho\hat{A}^\dagger\hat{A}\right)\   ,
\label{eq37}
\end{equation}
and $\kappa$, $\gamma$ and $\gamma_\phi$ represent  possible channels of dissipation representing, respectively, the cavity relaxation, qubit relaxation and qubit dephasing.
\medskip
\par

First, let us consider the effect due to each term of Eq.\ (\ref{master}). The Schr\"odinger term ($i[H,\rho]$)  neither creates nor destroys excitations, it can only transform a photon on an excitation on either qubit or transform an excitation on either qubit into a photon, the Lindblad terms can only destroy excitations or simply dephase the system. This means that, if we start with a system with a maximum value for $n$ total excitations, we need only to concern ourselves with matrix elements between states with $n$ or less total excitations, because neither term in the master equation is able to create excitations, only interchange them or destroy them. In particular, if we are interested in systems with  one excitation, we need only treat matrix elements between the states $\ket{0;00}\equiv\ket{0}$, $\ket{0;10}\equiv\ket{1}$, $\ket{0;01}\equiv\ket{2}$ and $\ket{1;00}\equiv\ket{3}$, where this relabeling was made for a cleaner notation. In this case, we can write the density matrix $\rho (t)$ as \cite{ballentine2014quantum}

\begin{equation}
\rho(t) = \sum_{i,j=0}^3\rho_{ij}(t)\ket{i}\bra{j} \label{rho} \   ,
\end{equation}
where the matrix elements obey the usual conditions, i.e., $\rho_{ij}=\rho_{ji}^*$, $0\leq\rho_{ii}\leq 1$ and $\sum_i\rho_{ii}=1$. Substituting (\ref{rho}) into (\ref{master}), we find

\begin{eqnarray}
\dot{\rho}_{ij} &=& -i\bra{i}[\hat{H},\rho]\ket{j}+\bra{i}\left[\kappa\mathcal{L}(\hat{a})\rho+\sum_{j=1}^2\left(\gamma\mathcal{L}(\sigma_j^-)\rho+\gamma_\phi\mathcal{L}\left(\sigma_j^{(3)}\right)\right)\right]\rho\ket{j}\nonumber
\nonumber\\
&=& {\dot{\rho}_S{_{ij}}}+\dot{\rho}_{\mathcal{L}_{ij}} \  ,
\label{Master2}
\end{eqnarray}
where

\begin{eqnarray}
\dot{\rho}_S{_{ij}} &=& -i\bra{i}[\hat{H},\rho]\ket{j}
\nonumber\\
\dot{\rho}_{\mathcal{L}_{ij}} &=&\bra{i}\left[\kappa\mathcal{L}(\hat{a})\rho+\sum_{j=1}^2\left(\gamma\mathcal{L}(\sigma_j^-)\rho+\gamma_\phi\mathcal{L}\left(\sigma_j^{(3)}\right)\right)\right]\rho\ket{j}. \label{rhodot}
\end{eqnarray}

\medskip
\par

In the low-temperature regime, we neglect the effect due to phonons on the properties of the graphene sheet and the decay of the system is governed by the cavity decay. In this regime, we have $\gamma=\gamma_\phi = 0$.  Equation (\ref{Master2}) gives rise to a set of ten independent, linear differential equations that are dealt with in Appendix for the case where the main form of dissipation is through cavity decay. In the Appendix, we focus our attention on the  steady states, which are preserved by the dynamics and on finding the differential equations that govern the dynamics for each element of the density matrix.  The non-trivial steady state is found in equations (\ref{SS1}) and (\ref{SS2}). The differential equation system (\ref{diffeq})  was then used to create a program that simulates the dynamics.
\medskip
\par

There are mainly two important results taken from the solution in Eq.\ (\ref{Master2}). The first one is that when dissipation comes mainly from cavity decay, meaning that at  low temperature $T$, the maximum entangled Bell state $\ket{0;\Psi_-}$,

\begin{equation}
\ket{0;\Psi_-} = \dfrac{1}{\sqrt{2}}\left(\ket{0;10}-\ket{0;01}\right) \label{Bell-}
\end{equation}
is a steady state and does not decay into the ground state. Another is that the characteristic decay time for all other states is of the order of $\tau= 1/\kappa$, as it should be expected.

\section{Concurrence as a Function of time}
\label{sec6}

\begin{figure}[H]
\begin{center}
\begin{subfigure}[b]{0.45\textwidth}
\includegraphics[width=\textwidth]{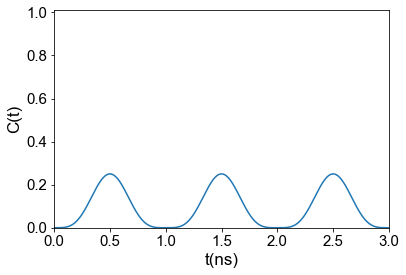}
\caption{}
\end{subfigure}
\hfill
\begin{subfigure}[b]{0.45\textwidth}
\includegraphics[width=\textwidth]{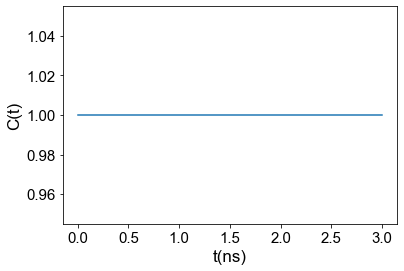}
\caption{}
\end{subfigure}
\\
\begin{subfigure}[b]{0.45\textwidth}
\includegraphics[width=\textwidth]{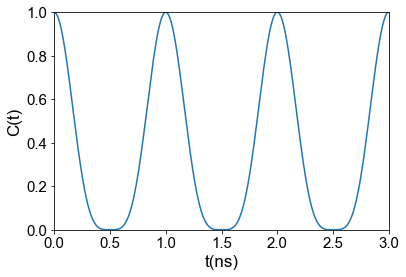}
\caption{}
\end{subfigure}
\hfill
\begin{subfigure}[b]{0.45\textwidth}
\includegraphics[width=\textwidth]{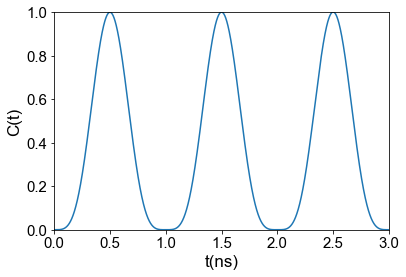}
\caption{}
\end{subfigure}
\end{center}
\caption{Concurrence between two qubits as a function of time for a system with no dissipation for four different initial states (a) $\rho_0 = \ket{0;10}\bra{0;10}$, meaning the first qubit starts in the excited state and the second in the ground state; (b) $\ket{0;\Psi_-}\bra{0;\Psi_-}$, the maximally entangled eigenstate of the unperturbed Hamiltonian (\ref{hami}); (c) $\ket{0;\Psi_+}\bra{0;\Psi_+}$, the maximally entangled state that is not an eigenstate of Eq.\ (\ref{hami}); and (d) $\ket{1;00}\bra{1;00}$, both qubits originally from the ground state and the cavity with a single photon.}
\label{FIG:1}
\end{figure}

As it was shown, the dynamics of a system that evolves under a master equation such as Eq.\  (\ref{Master2}) intertwines the states of both qubits. This means that one should expect the concurrence between those qubits to be a function of time.    At low temperature, the system can only decay through the loss of cavity photons. In this case, the only two steady states are the ground state, $\ket{g} = \ket{0;00}$, which has no concurrence ($C=0$) and the state $\ket{0;\Psi_-}$, given by Eq.\ (\ref{Bell-}).

\medskip
\par
A computer program that simulates the time evolution of a system governed by equation (\ref{diffeq}) was written and both the concurrence and probability for the system to decay to the ground state were calculated as a function of time. For all the simulations shown in this session, we chose the Rabi  frequency to be $\frac{2\pi}{2\sqrt{2}}$ns$^{-1}$.

We start with Fig.\  \ref{FIG:1} by showing the evolution of the concurrence as a function of time for a system with no dissipation ($\kappa = 0$). The effect of the dynamics is such that keeps the maximally entangled Bell state $\ket{0;\Psi_-}$ is constant, as expected, and creates oscillations between states $\ket{0;\Psi_+}$ and $\ket{1;00}$ with a frequency $\omega = 2\sqrt{2}\lambda$. The result is that the concurrence as a function of time presents oscillations with the same frequency between a maximum and minimum value that depends on the initial state.   However, the most interesting result  appears when we take dissipation into account. Figures  \ref{FIG:2}and \ref{FIG:3}  show what happens with such systems for two different initial states.

\begin{figure}[H]
\begin{center}
\begin{subfigure}[b]{0.3\textwidth}
\includegraphics[width=\textwidth]{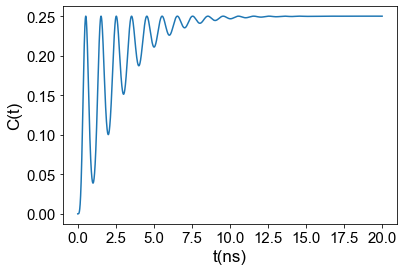}
\\
\includegraphics[width=\textwidth]{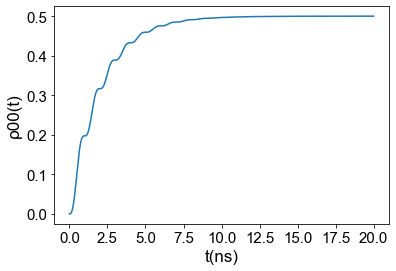}
\caption{}
\end{subfigure}
\hfill
\begin{subfigure}[b]{0.3\textwidth}
\includegraphics[width=\textwidth]{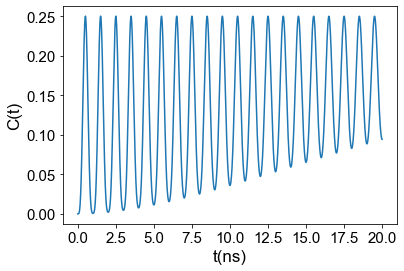}
\\
\includegraphics[width=\textwidth]{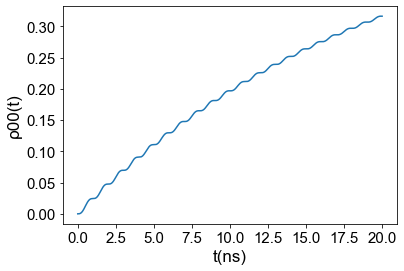}
\caption{}
\end{subfigure}
\hfill
\begin{subfigure}[b]{0.3\textwidth}
\includegraphics[width=\textwidth]{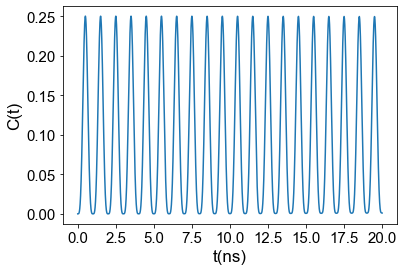}
\\
\includegraphics[width=\textwidth]{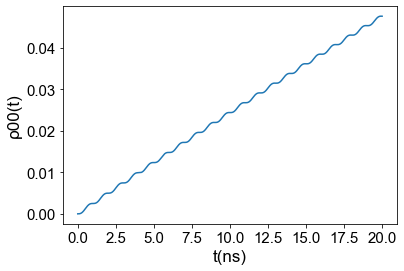}
\caption{}
\end{subfigure}
\end{center}
\caption{In the first row, concurrence between two qubits as a function of time for a system with dissipation for three different values of $\kappa$: (a) $\kappa = 1$ns$^{-1}$; (b) $\kappa = 0.1$ns$^{-1}$ and (c) $\kappa = 0.01$ns$^{-1}$. In the second row, we have the probability for the system to decay to the ground state as a function of time. In these simulation, the initial state of the system is always state $\rho_0 = \ket{0;10}\bra{0;10}$.}
\label{FIG:2}
\end{figure}

Figure \ref{FIG:2} shows how the concurrence evolves as a function of time for a system originally with one qubit in the excited state and one in the ground state for different cavity decay rates $\kappa$, while Fig.\ \ref{FIG:3}  shows the same information for a system originally in the maximally entangled Bell state $\ket{0;\Psi_+}$. The interesting thing to note is that although for the one in Fig.\ \ref{FIG:3}   the concurrence eventually decay to zero as the system decays to the ground state with probability one, for the initial state represented in Fig.\ \ref{FIG:2}, the decaying process actually increases the concurrence as time passes, and the decay probability goes to 0.5.
\medskip
\par
 \begin{equation}
\rho(t_0) = \rho_0\ket{\psi_0}\bra{\psi_0} + \rho_+\ket{\psi_+}\bra{\psi_+}+\rho_-\ket{\psi_-}\bra{\psi_-} + \sum_{i\neq j} \rho_{ij}\ket{\psi_i}\bra{\psi_j} + \rho_g\ket{0;00}\bra{0;00}  \  ,
\end{equation}
where, as in Eq (\ref{vec11}), $\ket{\psi_0} = \ket{0;\Psi_-}$ and the sum is taken over $i,j = +,-$ and $0$. The effect of the master equation on this basis is to keep $\rho_0$ constant while slowly of all other terms on the first manifold (all $\rho_{ij}$'s along with $\rho_\pm$) and increasing the probability for the system to be found on the ground state, $\rho_g$. The system will asymptotically decay to the final state.
\begin{figure}[H]
\begin{center}
\begin{subfigure}[b]{0.3\textwidth}
\includegraphics[width=\textwidth]{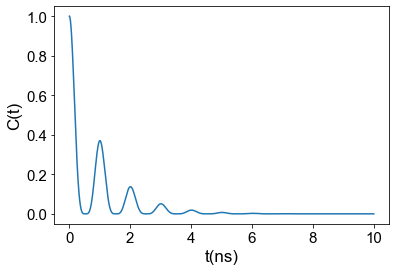}
\\
\includegraphics[width=\textwidth]{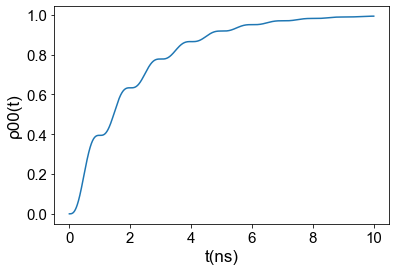}
\caption{}
\end{subfigure}
\hfill
\begin{subfigure}[b]{0.3\textwidth}
\includegraphics[width=\textwidth]{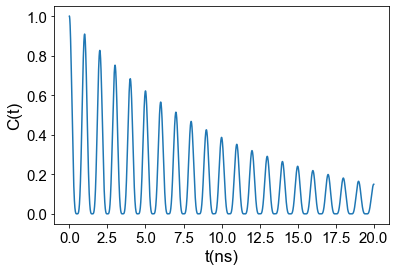}
\\
\includegraphics[width=\textwidth]{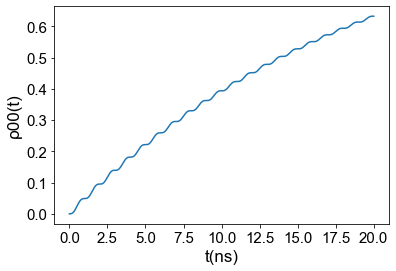}
\caption{}
\end{subfigure}
\hfill
\begin{subfigure}[b]{0.3\textwidth}
\includegraphics[width=\textwidth]{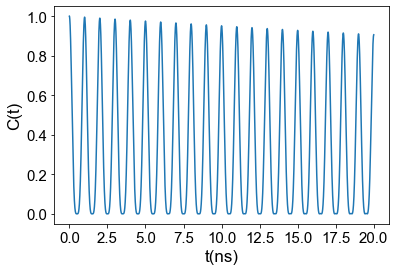}
\\
\includegraphics[width=\textwidth]{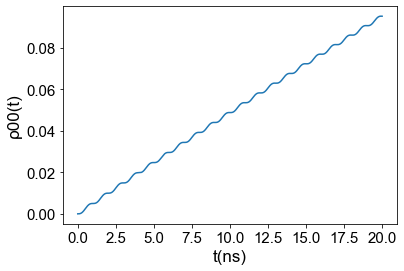}
\caption{}
\end{subfigure}
\end{center}
\caption{ The first row displays the concurrence between two qubits as a function of time for a system with dissipation for three different values of $\kappa$: (a) $\kappa = 1$ns$^{-1}$; (b) $\kappa = 0.1$ns$^{-1}$ and (c) $\kappa = 0.01$ns$^{-1}$. In the second row, we have the probability for the system to decay to the ground state as a function of time. In these simulations, the initial state of the system is always state $\rho_0 = \ket{0;\Psi_+}\bra{0;\Psi_+}$.}
\label{FIG:3}
\end{figure}

\begin{equation}
\rho(t\rightarrow \infty) = \rho_0\ket{\psi_0}\bra{\psi_0} + (1-\rho_0)\ket{0;00}\bra{0;00}  \   .
\label{rhof}
\end{equation}
When the system reaches the state represented in Eq.\ (\ref{rhof}), the concurrence can be calculated in a straightforward way by Eq.\ (\ref{concm}) and found to be

\begin{equation}
C(t\rightarrow \infty) = \rho_0^2,
\end{equation}
which is nothing more than the square of the probability for the system to be found in the maximally entangled Bell state $\ket{0;\Psi_-}$, defined in Eq.\ (\ref{Bell-}). The time which the system takes to decay to such a state is clearly dependent on the decay rate $\kappa$ of the cavity, but the end state depends only on the initial state. This means that, in some cases, concurrence can be protected from, \textit{and even created by}, dissipation in the system.

\section{concurrence between Excitons on a Graphene Sheet Under Strain}
\label{sec7}

We can now combine the results obtained in the previous sections to properly calculate the entanglement for our system of interest. In order to do so, we consider the same system as that in Ref.\ \cite{berman2020strain}, which is that for a graphene sheet under a strain-induced magnetic field of magnitude $\frac{B_z}{e} = 50 \ T$,   embedded in a GaAs microcavity with dielectric constant $\varepsilon = 13$. Assuming that the system is in resonance with the cavity, we need only to find the value of the Rabi splitting in order to calculate the time evolution of the concurrence between the excitons.

\medskip
\par

The Rabi splitting for excitons in graphene under a high magnetic field inside an optical microcavity has been calculated and found to be \cite{berman2009bose}

\begin{equation}
\hbar\lambda = e\left(\dfrac{\hbar\pi v_F r_B}{\sqrt{2}\epsilon W}\right)^{\frac{1}{2}}, \label{lambda}
\end{equation}
where $e$ is the electronic charge, $r_B=\sqrt{\hbar/B_z}$ is the magnetic length  for chosen magnetic field $B_z$, $\epsilon$ is the dielectric constant of the cavity, $v_F$ is the Fermi velocity for electrons in graphene and $W$ is the cavity's volume. The calculations in Ref.\  \cite{berman2009bose} that led to Eq.\ (\ref{lambda}) depended only on a single electron wave function and on the fact that the excitonic binding energy was much smaller than the band gap. The single electron wave function for an electron in a pseudomagnetic field is completely isomorphic to that of an electron in an external  magnetic field and the excitonic binding energy is also much smaller than the gap. This means that the Rabi splitting for excitons on a graphene sheet under strain is the same as the one for excitons on a graphene sheet in high magnetic field and is also given by Eq.\ (\ref{lambda}). Choosing the intensity of the strain-induced pseudomagnetic field as    $\frac{B_z}{e}=50 \ T$, a number that has been achieved experimentally for carbon nanotubes \cite{levy2010strain}, the Fermi velocity for electrons in graphene to be $v_F = 10^6 m/s$ and a GaAs microcavity ($\varepsilon = 13$) of volume $W=1.69 \times 10^3 \mu m^3$ \cite{vahala2003optical}, the Rabbi splitting becomes $\hbar \lambda = 0.027$ meV. The time evolution of the mean value of the concurrence was calculated for such a system that is inside a medium quality cavity, where photons have a medium lifetime of 1 $\mu $s ($\kappa = 10^{-3} ns^{-1}$). Those results are shown in Fig.\ \ref{FIG:4} for many different initial states.  The initial states $\ket{\Psi_0}$ considered in Fig.\  \ref{FIG:4} for $t=0$ could be prepared,  for example, by creating  the excitons by irradiating a precise femtosecond laser pump on the graphene sheet.

\begin{figure}[H]
\begin{center}
\begin{subfigure}[b]{0.3\textwidth}
\includegraphics[width=\textwidth]{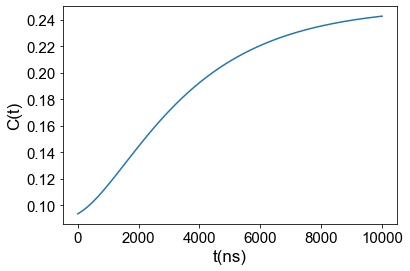}
\\
\includegraphics[width=\textwidth]{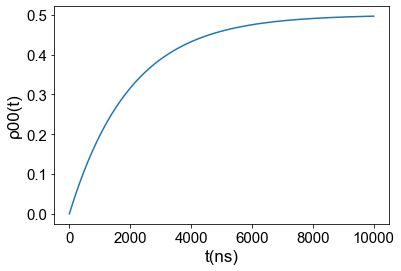}
\caption{}
\end{subfigure}
\hfill
\begin{subfigure}[b]{0.3\textwidth}
\includegraphics[width=\textwidth]{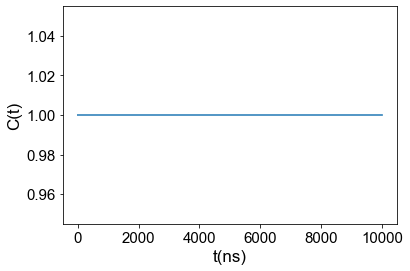}
\\
\includegraphics[width=\textwidth]{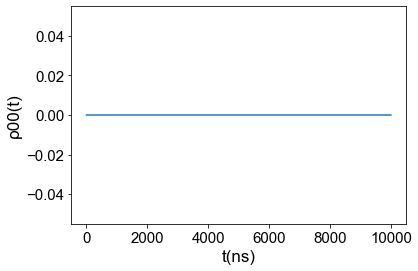}
\caption{}
\end{subfigure}
\hfill
\begin{subfigure}[b]{0.3\textwidth}
\includegraphics[width=\textwidth]{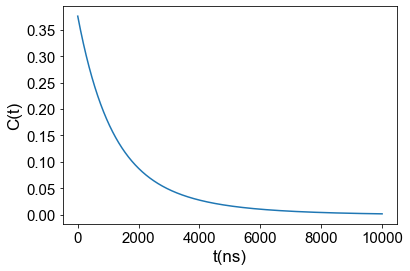}
\\
\includegraphics[width=\textwidth]{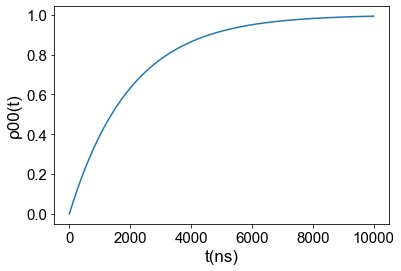}
\caption{}
\end{subfigure}
\end{center}
\caption{In the first row, the concurrence between a pair of qubits averaged in a single Rabi cycle  as a function of time for a system with dissipation for three chosen initial states $\ket{\Psi_0}$: (a) $\ket{\Psi_0} = \ket{0;10}$; (b) $\ket{\Psi_0} = \ket{0;\Psi_-}$ and (c) $\ket{\Psi_0} = \ket{0;\Psi_+}$.  The second row shows  the probability for the system  decaying to the ground state as a function of time.}
\label{FIG:4}
\end{figure}

Figure \ref{FIG:4}  shows that the effect due to cavity decay is not always the one which might seem intuitive, i.e.,   destroying the concurrence between excitons. This may be true if the system starts with no component in the Bell state $\ket{0;\Psi_-}$, when written in Bell's basis. However, if there are components on this basis vector, the final concurrence will not be zero and can  occasionally even increase with time due solely to the fact that the cavity decays. For such systems, the concurrence between two PMEs can be preserved throughout the entire lifetime of the excitons on graphene which, at low temperatures, can be a very long time since in this regime, excitons can only decay by photon emission and the $\ket{0;\Psi_-}$ state is protected from such decay. Even though we considered excitons on a graphene sheet under strain throughout this paper, the same results apply to any excitonic system in which excitons have a set of discrete energy levels. One such example is for trapped excitons on a TMDC monolayer or bilayer. One can make such a trap by, for example, pinning the TMDC with a thin needle \cite{negoita1999stretching} or by applying a potential difference between two layers at a given point \cite{voros2006trapping}.

\section{Possible Physical Realizations}
\label{real}

The following physical realizations are possible for similar qubits based on intervalley excitons:   First, for pristine (or gapless) graphene with donor impurities or due to the control electrode, the chemical potential will be in the upper (conductivity) band. Then a magnetic transition is possible between Landau levels in the same conduction band. This transition requires pumping with terahertz photons. As a result of pumping by terahertz photons with linear polarization, intervalley excitons will arise.  Secondly, in pristine graphene, pumping by photons in the optical region of the spectrum is necessary, which transfers electrons from the  filled Landau level in the valence band to an unfilled Landau level in the conduction band.

\medskip
\par
Note that the results of this work are valid not only for inter-valley excitons in graphene in a pseudomagnetic field, but also for excitons in other physical systems with a discrete spectrum. For example, one can use excitons in a TMDC material in a 2D trap created by the deformation potential from the tip of a scanning probe microscope. Another option is a system of Frenkel excitons in 2D organic materials. For efficient control of such qubits and their use in quantum technologies, these systems can be placed in a spatially limited optical cavity with a discrete photon spectrum with resonant frequencies for qubits.

\medskip
\par
 The most important property of qubits used in quantum technologies is quantum entanglement, and no entanglement occurs during adiabatic changes in the system of qubits. For entanglement, a non-adiabatic effect is necessary. With such an effect, interesting phenomena also arise in quantum electrodynamics in a cavity such as the dynamic Lamb effect  (see Refs.~\cite{berman2016quantum,klucevsek2017intersection} and references therein). In fact, quantum entanglement should appear immediately upon the above-described appearance of intervalley excitons after pumping with linearly polarized light. Similarly, quantum entanglement of qubits should appear immediately after the pumping of excitons in the trap or pumping of Frenkel  excitons. Calculating the properties of such quantum entanglement and its time dependence is the goal of our work. Modern quantum technologies and the prospects for their development are based on the use of quantum two-level systems, i.e., qubits.

\medskip
\par

A number of physical realizations of qubits in systems with discrete spectra have been mentioned.   These include  ultra cold ions and atoms in traps, impurities in diamond,  and various types of superconducting qubits. Ordinary extended semiconductors and new 2D materials, TMDC, have a band spectrum. But, in a transverse magnetic fields 2D systems have a discrete spectrum consisting of degenerate Landau levels. However, when electrons are excited from filled Landau  levels to unfilled ones, electrons and holes due to the Coulomb attraction form 2D magnetoexcitons, the energy of which depends continuously on the integral of motion in a magnetic field, i.e.,  the magnetic momentum. This integral of motion is a consequence of the invariance of the system with respect to simultaneous translation and gauge transformation (see  Refs.~\cite{lozovik1997magnetoexcitons,lozovik1997magnetoexciton2}). As a result of the continuous dependence of the magnetoexciton energy on the magnetic pulse, the full spectrum of the system is not discrete, but consists of bands (see Ref.~\cite{lerner1980mott}).

\medskip
\par
Principally new properties of 2D systems arise in pseudomagnetic fields, which arise, in particular,   in graphene under strain. This is due to the fact that the effect of pseudomagnetic fields does not depend on the sign of the charge and is the same for electrons and holes.Another important property of graphene is the chirality of only two independent valleys, which leads to the fact that circularly polarized light is absorbed in different valleys of graphene, depending on the sign of circular polarization.  Additionally, the consequence of chirality is that the resulting pseudomagnetic field during deformation has a different sign in different valleys.  Consequently, the appearance of a pseudomagnetic field due to  deformation of graphene does not contradict the absence of violation of invariance with respect to time reversal contrary to the case of graphene in real magnetic field.
 
\medskip
\par
If graphene is pumped by plane polarized photons, they are absorbed in both valleys, accompanied by the appearance of electrons and holes in them.  As a result of the Coulomb attraction, these electrons and holes form excitons not only in the same valley but also from different valleys. They are more stable due to forbidden luminescence  as a consequence of the law of energy and momentum conservation.  It can be easily shown that action of pseudomagnetic fields, contrary to real  magnetic fields, does not depend on charges.  Also, pseudomagnetic fields in different valleys have opposite sign and act the same on the electron and the hole.  Consequently,   intervalley  pseudomagnetoexcitons have discrete spectra and  can be used as quantum elements in quantum technologies.

\section{Concluding Remarks}
\label{sec8}

In this paper, we have investigated quantum entanglement between a pair of excitons, formed by an electron and a hole from different valleys in a graphene monolayer under strain within an optical microcavity.  We first developed a Jaynes-Cummings like model for two qubits coupled to a single cavity mode that governs the system in the rotating wave approximation (RWA). We then calculated the energy eigenstates of such a model. We measured the degree of entanglement of such states by calculating the concurrence between the two qubits in each of these eigenstates. It was shown that if the system is allowed to decay only through the emission of cavity photons, which is the case at low temperatures, there is a maximally entangled eigenstate which is protected from decay. We showed that the existence of such a state leads to the counter-intuitive consequence that, for some initial states of the system, the fact that the cavity is leaky can actually lead to an increase in the average concurrence on the timescales of the average photonic lifetime. Lastly, we used the energy eigenfunctions  and eigenvalues for PMEs obtained in Ref.\ \cite{berman2020strain} to formally calculate the time evolution of the concurrence between two PMEs in graphene.  In addition, we have discussed the  applicability of our approach for different 2D materials, such as TMDC monolayers.

\acknowledgments

O.L.B. is grateful to  R.~Ya.
Kezerashvili and G.~V. Kolmakov for the valuable discussions. The
authors are grateful for support by grants: O.L.B. - the PSC CUNY
Grant No. 61538-00 49; O.L.B. and G.P.M. and U.S. ARO grant No.
W911NF1810433. GG acknowledges the support from the US AFRL Grant No. FA9453-21-1-0046. The part devoted to new physical realizations of qubits  and ququarts was supported by the RFBR grants No. 20-
02-00410 and No. 20-52-00035. Yu.E.L. acknowledged Basic Research Program at the National Research University HSE.

\bibliographystyle{apsrev4-2}
\bibliography{bib}

\appendix

\section{Dissipative dynamics}
\label{sec9}

We determine the solution for the set of differential equations represented by Eq.\ (\ref{Master2}), in the case where only the cavity decay is included ($\gamma = \gamma_\phi - 0$). For this, we first turn our attention to the dissipative term represented by Eq.\ (\ref{rhodot}). If $\gamma = \gamma_\phi - 0$, this equation becomes

\begin{eqnarray}
\dot{\rho}_{\mathcal{L}_{ij}} &=& \bra{i}\kappa\mathcal{L}(\hat{a})\rho\ket{j}\nonumber\\
&=&\hat{a}\rho\hat{a}^\dagger -\dfrac{1}{2}\left(\hat{a}^\dagger\hat{a}\rho+\rho\hat{a}^\dagger\hat{a}\right)\label{dissip}.
\end{eqnarray}

\medskip
\par

On this manifold, the effect of the operator $\hat{a}$ is to take the system from the state with one photon and zero excitonic excitations,  $\ket{3}= \ket{1;00}$, to the ground state $\ket{0}=\ket{0;00}$,  i.e.,

\begin{eqnarray}
\hat{a} &=& \ket{0}\bra{3}\label{a} \\
\hat{a}^\dagger &=& \ket{3}\bra{0}\label{ad}.
\end{eqnarray}
Making use of Eq.] (\ref{a}-\ref{ad}) in Eq.  \  (\ref{dissip}), we find

\begin{equation}
\mathcal{L}(\hat{a})\rho = \rho_{33}\left(\ket{0}\bra{0}-\ket{3}\bra{3}\right) -\dfrac{1}{2}\sum_{j=0}^2\left(\rho_{3j}\ket{3}\bra{j}+\rho_{j3}\ket{j}\bra{3}\right)    \  ,
\end{equation}
where $\rho_{ij}=\bra{i}\rho\ket{j}$.  This leads to

\begin{eqnarray}
\dot{\rho}_{\mathcal{L}_{00}} &=& \kappa\rho_{33}\label{firstL}
\nonumber\\
\dot{\rho}_{\mathcal{L}_{33}} &=& -\kappa\rho_{33}
\nonumber\\
\dot{\rho}_{\mathcal{L}_{i3}} &=& -\dfrac{\kappa}{2}\rho_{i3} ; i=0,1,2 \nonumber\\
\dot{\rho}_{\mathcal{L}_{01}}=\dot{\rho}_{\mathcal{L}_{02}} = \dot{\rho}_{\mathcal{L}_{11}} &=& \dot{\rho}_{\mathcal{L}_{12}} = \dot{\rho}_{\mathcal{L}_{22}} = 0 \   .
\label{lastL} ,
\end{eqnarray}
All the other coefficients are obtained by recalling  that $\rho_{ij}=\rho_{ji}^\ast$.

\medskip
\par

In order to find the terms $\dot{\rho}_{S_{ij}}$, we first determine the result obtained when  the Hamiltonian acts on each state of our basis, i.e.,

\begin{eqnarray}
\hat{H}\ket{0} &=& 0
\nonumber\\
\hat{H}\ket{1} &=& \omega_0\ket{1}+\lambda\ket{3}
\nonumber\\
\hat{H}\ket{2} &=& \omega_0\ket{2}+\lambda\ket{3}
\nonumber\\
\hat{H}\ket{3} &=& \omega_k\ket{3}+\lambda\left(\ket{1}+\ket{2}\right)  \  .
\end{eqnarray}
Making use of this result in Eq. (\ref{eq37}), we find

\begin{eqnarray}
\dot{\rho}_{S_{00}} &=& 0\label{firstS}
\nonumber\\
\dot{\rho}_{S_{01}} &=& i\omega_0\rho_{01}+i\lambda \rho_{03}
\nonumber\\
\dot{\rho}_{S_{02}} &=& i\omega_0\rho_{02}+i\lambda \rho_{03}
\nonumber\\
\dot{\rho}_{S_{03}} &=& i\omega_k\rho_{03}+i\lambda( \rho_{01}+\rho_{02})
\nonumber\\
\dot{\rho}_{S_{11}} &=& -i\lambda (\rho_{31}-\rho_{13})
\nonumber\\
\dot{\rho}_{S_{12}} &=& -i\lambda (\rho_{32}-\rho_{13})
\nonumber\\
\dot{\rho}_{S_{13}} &=&-i(\omega_0-\omega_k)\rho_{13} -i\lambda (\rho_{33}-\rho_{11}-\rho_{12})
\nonumber\\
\dot{\rho}_{S_{22}} &=& -i\lambda (\rho_{32}-\rho_{23})
\nonumber\\
\dot{\rho}_{S_{23}} &=&-i(\omega_0-\omega_k)\rho_{23} -i\lambda (\rho_{33}-\rho_{22}-\rho_{21})
\nonumber\\
\dot{\rho}_{S_{33}} &=& -i\lambda (\rho_{13}-\rho_{31}+\rho_{23}-\rho_{32}) \label{lastS}.
\end{eqnarray}
By combining Eqs.\ (\ref{firstL})  through (\ref{lastL}), with Eqs.\ (\ref{firstS})  through (\ref{lastS}) and using the fact that $\rho_{ij} = \rho_{ji}^\ast$, we can rewrite Eq.\ (\ref{Master2}) as

\begin{equation}
\dot{\rho}_V = \left[i\Lambda+\frac{\kappa}{2}\Gamma\right]\rho_V ,
\end{equation}
where $\rho_V$ is the column vector

\begin{equation}
\rho_V = (\rho_{00},\rho_{01},\rho_{10},\rho_{02},\rho_{20},\rho_{03},\rho_{30},\rho_{11},\rho_{12},\rho_{21},\rho_{13},\rho_{31},\rho_{22},\rho_{23},\rho_{32},\rho_{33},)^T,
\end{equation}
the matrix $\Lambda$ is given by

\begin{equation}
\Lambda = \begin{pmatrix}
0&0&0&0&0&0&0&0&0&0&0&0&0&0&0&0\\
0&\omega_0&0&0&0&\lambda&0&0&0&0&0&0&0&0&0&0\\
0&0&-\omega_0&0&0&0&-\lambda&0&0&0&0&0&0&0&0&0\\
0&0&0&\omega_0&0&\lambda&0&0&0&0&0&0&0&0&0&0\\
0&0&0&0&-\omega_0&0&-\lambda&0&0&0&0&0&0&0&0&0\\
0&\lambda&0&\lambda&0&\omega_k&0&0&0&0&0&0&0&0&0&0\\
0&0&-\lambda&0&-\lambda&0&-\omega_k&0&0&0&0&0&0&0&0&0\\
0&0&0&0&0&0&0&0&0&0&\lambda&-\lambda&0&0&0&0\\
0&0&0&0&0&0&0&0&0&0&\lambda&0&0&0&-\lambda&0\\
0&0&0&0&0&0&0&0&0&0&0&-\lambda&0&\lambda&0&0\\
0&0&0&0&0&0&0&\lambda&\lambda&0&(\omega_k-\omega_0)&0&0&0&0&-\lambda\\
0&0&0&0&0&0&0&-\lambda&0&-\lambda&0&(\omega_0-\omega_k)&0&0&0&\lambda\\
0&0&0&0&0&0&0&0&0&0&0&0&0&\lambda&-\lambda&0\\
0&0&0&0&0&0&0&0&0&\lambda&0&0&\lambda&(\omega_k-\omega_0)&0&-\lambda\\
0&0&0&0&0&0&0&0&-\lambda&0&0&0&-\lambda&0&(\omega_0-\omega_k)&\lambda\\
0&0&0&0&0&0&0&0&0&0&-\lambda&\lambda&0&-\lambda&\lambda&0
\end{pmatrix},
\end{equation}
and the matrix $\Gamma$ is given by
\begin{equation}
\Gamma = \begin{pmatrix}
0&0&0&0&0&0&0&0&0&0&0&0&0&0&0&2\\
0&0&0&0&0&0&0&0&0&0&0&0&0&0&0&0\\
0&0&0&0&0&0&0&0&0&0&0&0&0&0&0&0\\
0&0&0&0&0&0&0&0&0&0&0&0&0&0&0&0\\
0&0&0&0&0&0&0&0&0&0&0&0&0&0&0&0\\
0&0&0&0&0&-1&0&0&0&0&0&0&0&0&0&0\\
0&0&0&0&0&0&-1&0&0&0&0&0&0&0&0&0\\
0&0&0&0&0&0&0&0&0&0&0&0&0&0&0&0\\
0&0&0&0&0&0&0&0&0&0&0&0&0&0&0&0\\
0&0&0&0&0&0&0&0&0&0&0&0&0&0&0&0\\
0&0&0&0&0&0&0&0&0&0&-1&0&0&0&0&0\\
0&0&0&0&0&0&0&0&0&0&0&-1&0&0&0&0\\
0&0&0&0&0&0&0&0&0&0&0&0&0&0&0&0\\
0&0&0&0&0&0&0&0&0&0&0&0&0&-1&0&0\\
0&0&0&0&0&0&0&0&0&0&0&0&0&0&-1&0\\
0&0&0&0&0&0&0&0&0&0&0&0&0&0&0&-2
\end{pmatrix}.
\end{equation}

First, we note that, as expected, $\dot{\rho}_{00} = - (\dot{\rho}_{11}+\dot{\rho}_{22}+\dot{\rho}_{33})$, which means that the equation for $\rho_{00}$ is redundant and is not needed to solve the system, so we can eliminate the first entry to $\rho_V$ and the first line and column of the matrices $\Lambda$ and $\Gamma$, without any effect. Then, we realize that the remaining problem can be expressed as

\begin{equation}
\dot{\rho}_V = M\rho_V,
\end{equation}
where the matrix $M$ can be written as
\begin{equation}
M = \begin{pmatrix}
B & 0_{6\times 9} \\
0_{9\times 6} & A
\end{pmatrix},
\end{equation}
where $B$ is a $6\times 6$ matrix, $A$ is a $9\times 9$ matrix and $0_{m\times n}$ is a $m\times n$ matrix with all-zero elements. This means that the time evolution does not mix elements $\rho_{0i}$ with $i\neq 0$ with elements $\rho_{ij}$ with $i,j\neq 0$, nor does it mix $\rho_{00}$ with $\rho_{0i}$ with $i\neq 0$. Therefore, if $\ket{0}\rho\bra{i}=0$ for $i=1,2,3$ at $t=0$, $\ket{0}\rho\bra{i}=0$. These are the cases we are interested in, since we want to see how the Hamiltonian eigenvalues $\ket{\psi_0}$ and $\ket{\psi_\pm}$ evolve in time, and they do not involve the state $\ket{0;00}$.

\medskip
\par
For such cases, the time evolution is given by

\begin{equation}
\dot{\overline{\rho}} = A\overline{\rho}\label{diffeq}
\end{equation}
with
\begin{equation}
\overline{\rho} = (\rho_{11},\rho_{12},\rho_{21},\rho_{13},\rho_{31},\rho_{22},\rho_{23},\rho_{32},\rho_{33})^T \label{string}
\end{equation}
 and

 \begin{equation}
 A = \begin{pmatrix}
 0&0&0&i\lambda&-i\lambda&0&0&0&0\\
 0&0&0&i\lambda&0&0&0&-i\lambda&0\\
 0&0&0&0&-i\lambda&0&i\lambda&0&0\\
 i\lambda&i\lambda&0&a&0&0&0&0&-i\lambda\\
 -i\lambda&0&-i\lambda&0&a^*&0&0&0&i\lambda\\
 0&0&0&0&0&0&i\lambda&-i\lambda&0\\
 0&0&i\lambda&0&0&i\lambda&a&0&-i\lambda\\
 0&-i\lambda&0&0&0&-i\lambda&0&a^*&i\lambda\\
 0&0&0&-i\lambda&i\lambda&0&-i\lambda&i\lambda&-\kappa
 \end{pmatrix},
 \end{equation}
 where $a=-\frac{\kappa}{2}+i(\omega_k-\omega_0)$. Now, all that needs to be done is to find the eigenvectors of this matrix and select from these the ones in which $\rho_{ij}=\rho_{ji}^\ast$, which will be the ones with meaningful physical reality.
\medskip
\par

The spectral decomposition of $A$ is rather complicated with very long expressions for most of the eigenvectors and eigenvalues. However, there is one interesting solution that arises with an eigenvalue of zero, which is represented by the vector

\begin{equation}
\overline{\rho}_0 = (1,-1-1,0,0,1,0,0,0)^T,
\end{equation}
which means that, by going back to the definition of the string $\overline{\rho}$ from Eq.\ (\ref{string}) and using Eq.\ (\ref{master}), the state represented by the density operator $\rho$ given by

\begin{equation}
\rho_{ST} = \dfrac{1}{2}\left(\ket{1}\bra{1}+\ket{2}\bra{2}-\ket{1}\bra{2}-\ket{2}\bra{1}\right)\label{Bell}
\end{equation}
is a stationary state. It is a simple matter to show that Eq.\ (\ref{Bell}) canbe expressed as

\begin{equation}
\rho_{ST} = \ket{\Psi^-}\bra{\Psi^-}\label{SS1}
\end{equation}
where the maximum entangled Bell state $\ket{\Psi^-}$ is given by \cite{ballentine2014quantum}

\begin{equation}
\ket{\Psi^-} = \dfrac{1}{\sqrt{2}}\left(\ket{1}-\ket{2}\right)=\dfrac{1}{\sqrt{2}}\left(\ket{0;10}-\ket{0;01}\right).\label{SS2}
\end{equation}
It is important to note that all other eigenvalues have a negative real part, which means that all other eigenvectors represent states that decay to the ground state exponentially with time. In the case of resonance ($\omega_0=\omega_k$), the eigenvalues of $A$ are $\lambda_1=0$, $\lambda_2=\lambda_3 = -\dfrac{\kappa}{2}$, $\lambda_4=\lambda_5=-\dfrac{\kappa}{2}-i\sqrt{32\lambda^2-\kappa^2}$ and $\lambda_6=\lambda_7=\lambda_8=\lambda_9=-\dfrac{\kappa}{4}+i\sqrt{32\lambda^2-\kappa^2}$, which means that all states other then $\rho_{ST}$ decay with a characteristic time no larger than $\tau = \dfrac{4}{\kappa}$.

\end{document}